# Development of a 3-D Position Sensitive Neutron Detector Based on Organic Scintillators with Double Side SiPM Readout

Yang TIAN[1,2,*], Yidong FU[1,2], Yulan LI[1,2], Yuanjing LI[1,2]

1. Dept. of Engineering Physics, Tsinghua University, Beijing 100084, China

2. Key Laboratory of Particle & Radiation Imaging (Tsinghua University), Ministry of Education, Beijing 100084, China

**Abstract** A 3-D position sensitive neutron detector is being developed based on a plastic scintillator array. A double side SiPM readout is used to determine the depth of interaction (DOI) in each scintillator unit. In the preliminary test, the DOI in a 254 × 6 × 6 mm$^3$ SP101 plastic scintillator is measured at different positions using a collimated $^{60}$Co source. The SiPM (KETEK PM6660) signals are recorded by a 2.5 GS/s digital oscilloscope. The DOI results are calculated using both the amplitude and the temporal information. Position resolutions (FWHM) of 2.5 cm and 6.6 cm are realized, respectively. A detector based on a 2-D array is capable of recording the 3-D position information of the incident neutron. The 3-D detector is to be used together with a neutron time projection chamber as a directional fast neutron detector. According to the simulation results, the angular resolution (8 degree FWHM) is much better than that of a typical neutron scatter camera.

## 1. Introduction

The most common method of fast neutron detection is based on elastic scattering of neutrons by light nuclei. The kinematics of neutron elastic scattering is the basis of several directional fast

[*] Corresponding author. E-mail address: yangt@mail.tsinghua.edu.cn (Yang TIAN).

neutron detectors (Fig. 1) developed in the past decade [1]-[3]

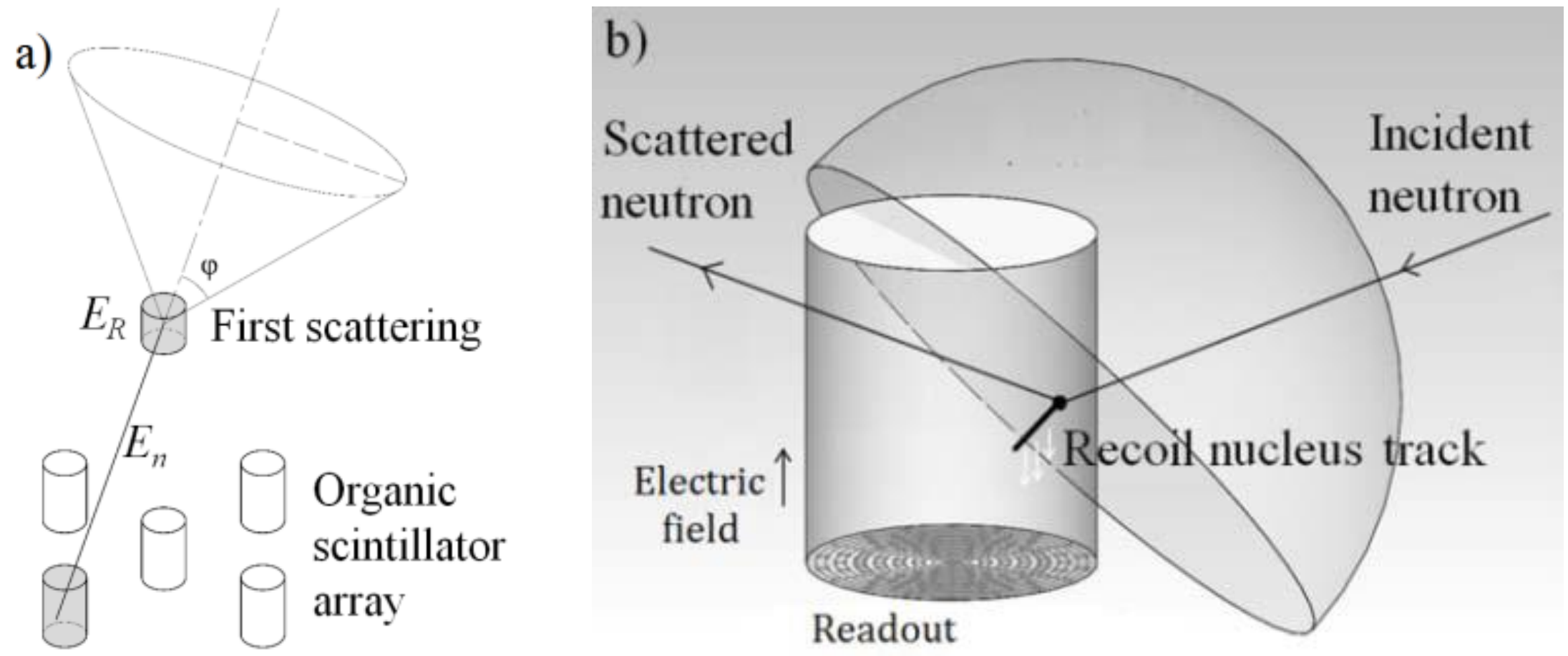

Fig. 1 Directional fast neutron detectors. a) a neutron scatter camera; b) a neutron TPC.

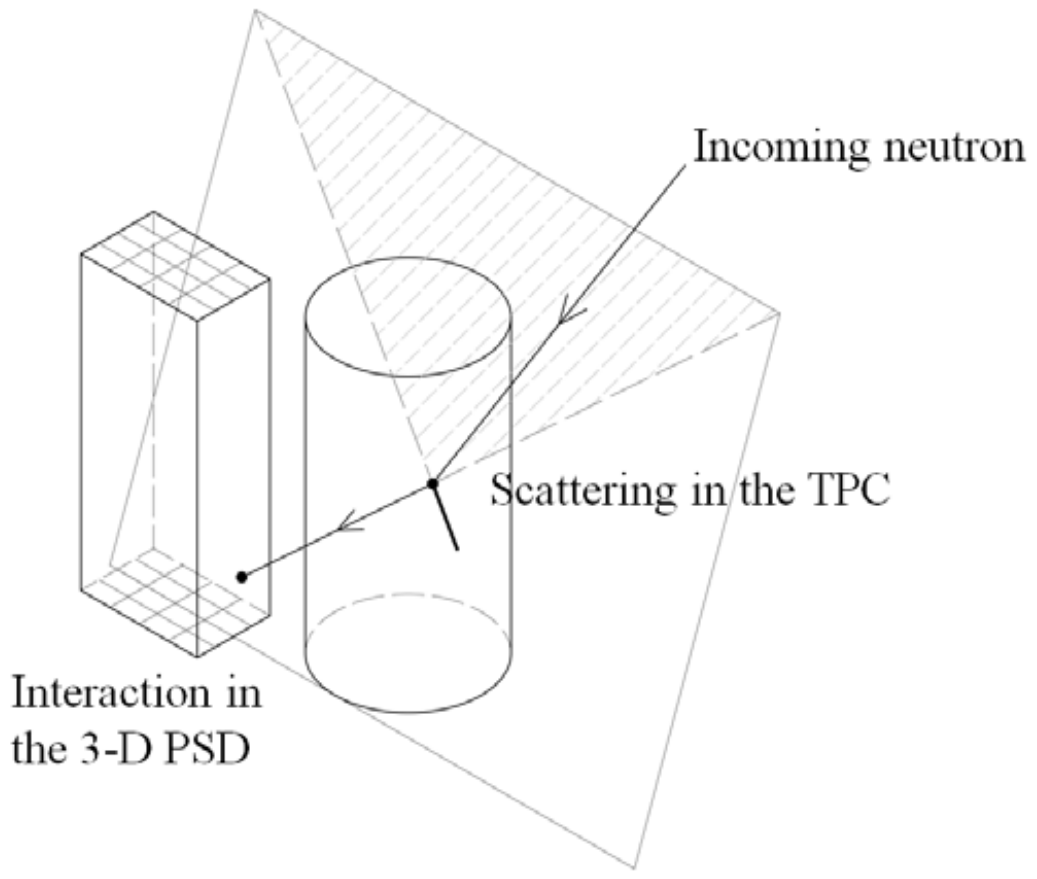

Fig. 2 Directional fast neutron detection method based on a neutron TPC and a 3-D PSD.

We propose a new directional fast neutron detection method based on a neutron time projection chamber (TPC) and a 3-D position sensitive neutron detector (PSD) (Fig. 2). The neutron TPC is a omnidirectional system, i.e., a 4π field of view. The TPC alone can provide a quick but rough detection to indicate the directions of interest. A fine measurement can be done by using the coincidence events recorded by both the TPC and the 3-D PSD. The direction of the incident neutron can be limited within one quadrant of a plane (the dashed area in Fig. 2). Both the track and the energy of the recoil proton can be measured by the TPC, so there is no need to measure the energy of the scattered neutron. According to the simulation results, the angular resolution can be better than 8 degree (FWHM) using the simple back projection method (Fig. 3).

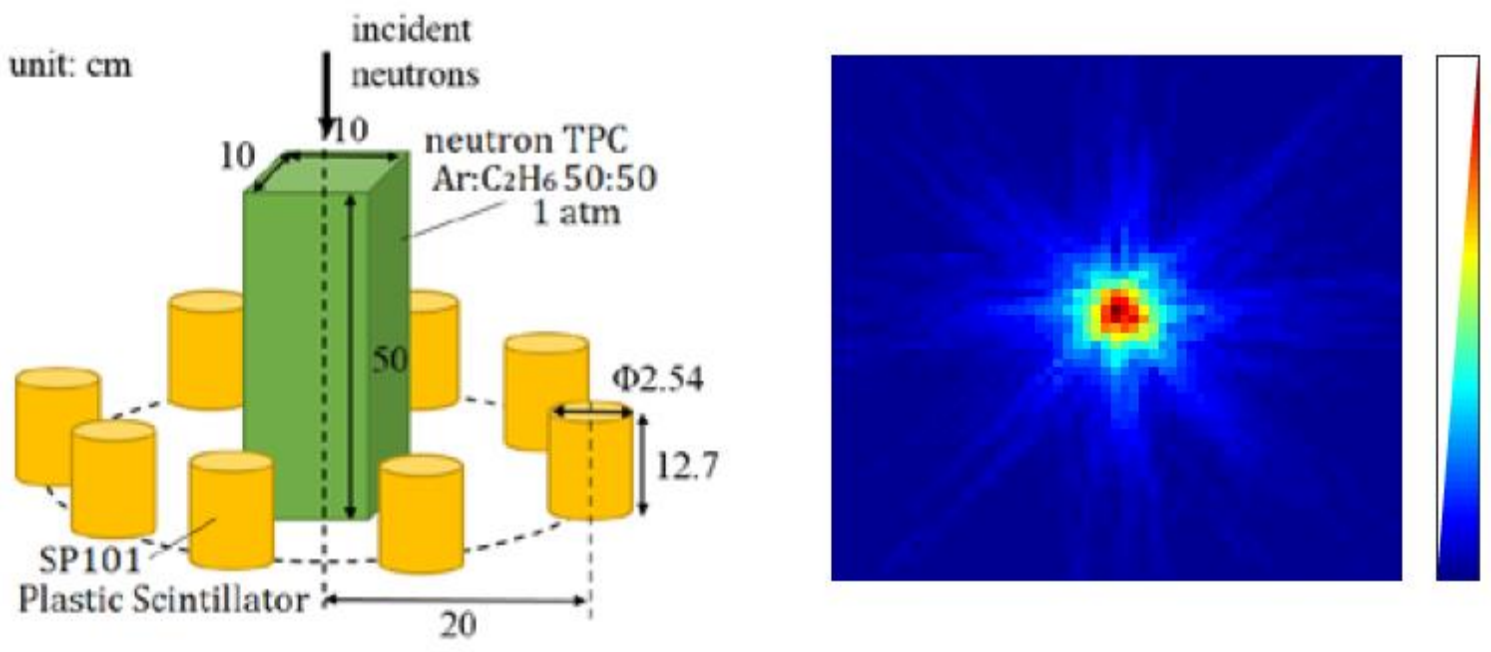

Fig. 3 Simulation setup and results of a neutron imaging TPC.

To guarantee a reasonable detection efficiency, 2-D organic scintillator array of large volume is used in the design. A 3-D PSD can be realized if the depth of interaction (DOI) is measured using double side readout.

The recent development of the silicon photomultiplier (SiPM) greatly facilitate the design of the 2-D photo detector and the double side readout. A plastic scintillator with double side SiPM readout (i.e., one basic unit of the scintillator array) is tested with different electronics. The preliminary results are presented in this paper.

## 2. Experimental setup

One basic unit of the scintillator with double side readout is illustrated in Fig. 4. A 254 × 6 × 6 $mm^3$ SP101 plastic scintillator is coupled to two KETEK PM6660 SiPMs (6 × 6 $mm^2$ effective area) and wrapped by the PTFE tapes. The SP101 scintillator has a light yield of approximately 8000 photons/MeV and a decay time of 2.8 ns. The signals are recorded directly by a 2.5 GS/s digital oscilloscope (LeCroy HDO6104). A collimated $^{60}$Co source is used to test the spatial resolution.

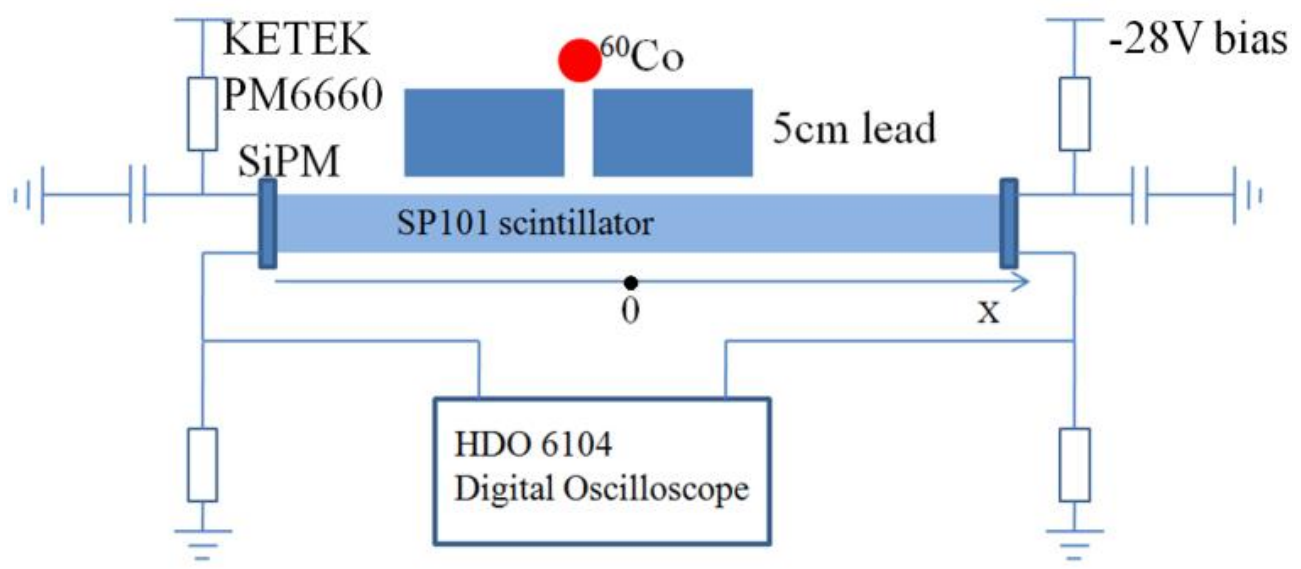

Fig. 4 The experimental setup. The resistor connected to the oscilloscope stands for the input impendence of the readout circuit.

Two different methods are used to calculate the DOI. One is based on the signal amplitudes. The DOI can be determined by (1), where α is the light attenuation coefficient [4]. $E_2$ and $E_1$ are the signal amplitudes at the two ends. They can be measured by using charge sensitive preamplifiers or integrating the fast current signals. In the other method, the DOI is calculated by the time difference. Two different readout circuits are fabricated accordingly. The schematics are shown in Fig. 5.

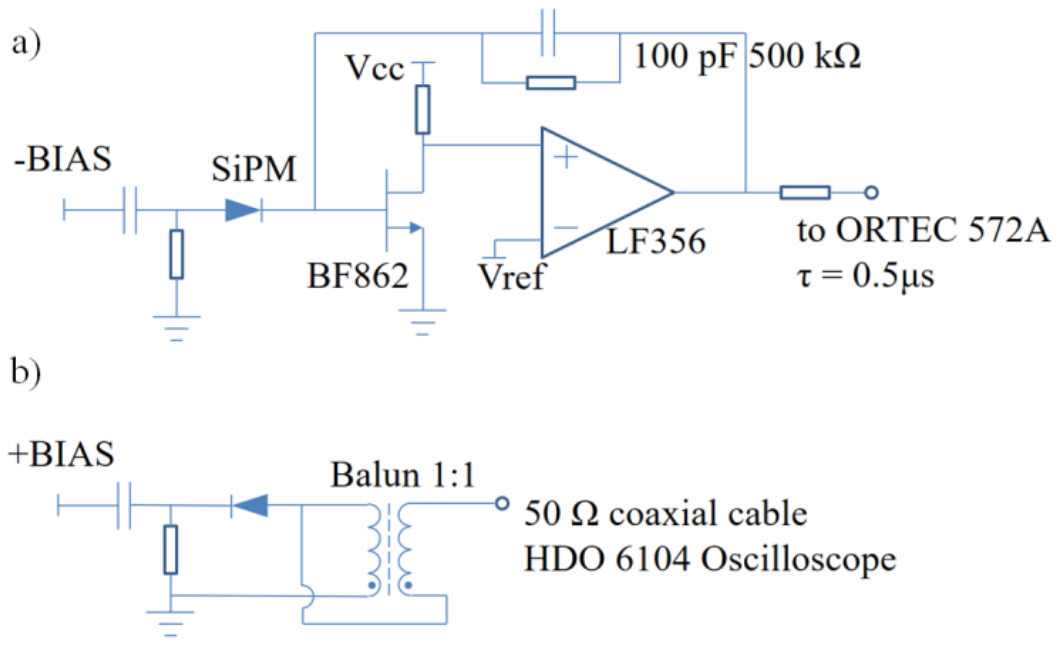

Fig. 5 The schematics of the readout circuits. a) a charge sensitive preamplifier for the amplitude based method; b) a fast balun for time difference based method. The latter can also be used to estimate the number of photons collected at both ends (i.e., amplitude method) by integrating the fast signal within a 300 ns time window.

## 3. Results

The relationship between the measured signal amplitudes and the position is shown in Fig. 6a. The result agree well with (1). The value of αL is 2.1 (L = 254 mm). The energy spectrum can also be derived from the signal amplitudes (Fig. 6b). However, the energy calibration is difficult without a clear Compton edge. The spectrum is simply divided into 4 ranges. Spatial resolutions within different ranges are shown in Fig. 6c.

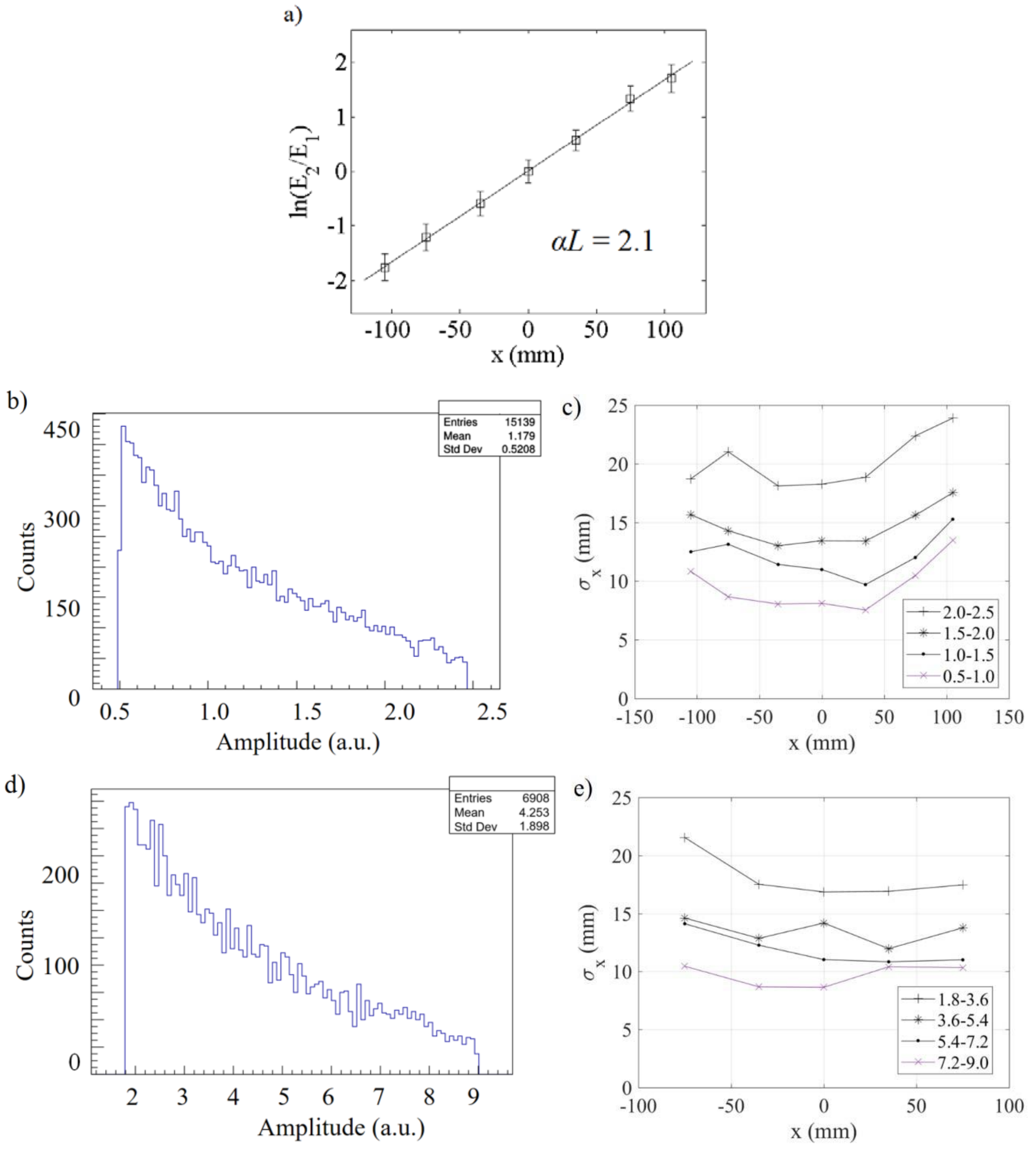

Fig. 6 Experimental results of the amplitude based method. a) the linear fit according to (1); b) the derived energy spectrum using the circuit in Fig. 5a; c) spatial resolution within different energy ranges in b); d) the derived energy spectrum using the circuit in Fig. 5b; e) spatial resolution within different energy ranges in d).

Similar measurements and calculations were carried out using the balun readout. The amplitude was calculated by integrating the signal within 300 ns time window). The results are shown in Fig. 6d. The time difference was measured using the rise edge of the balun signal at 50% trigger level. The effective light propagation speed is estimated to be 10.3 cm/ns.

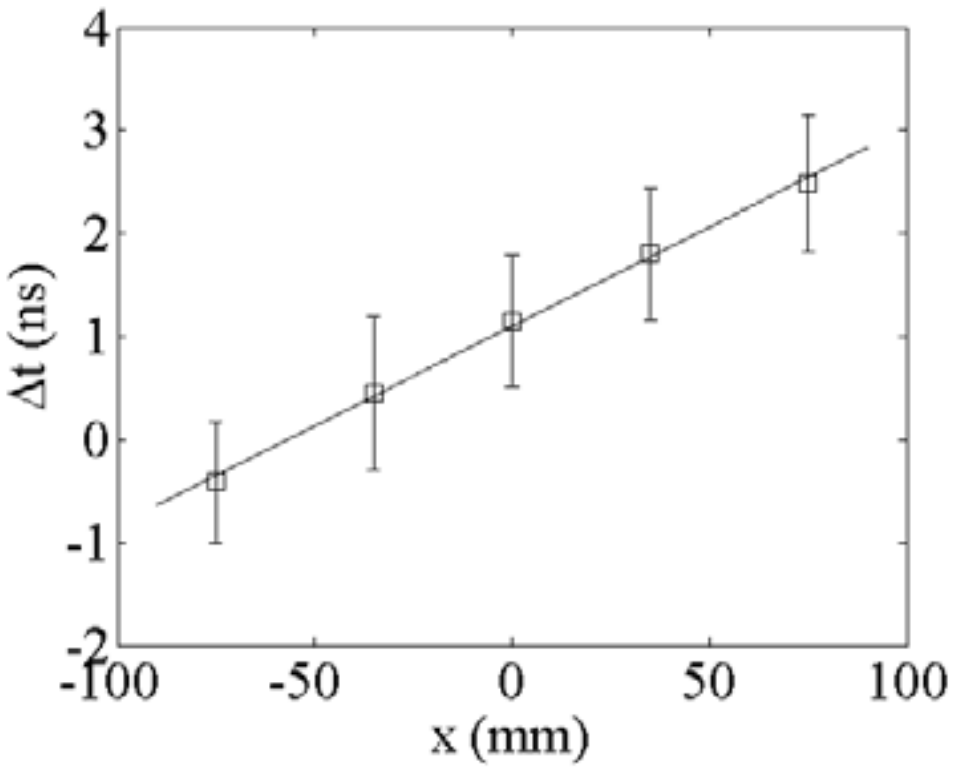

Fig. 7 Experimental results of the time difference based method.

## 3. Discussion and future work

The amplitude based method shows much better spatial resolution. The CSP and balun circuits give similar results. As the optimal value of αL is 2.9 [4], the spatial resolution can still be improved. The measured sigma of the time difference is around 650 ps, which corresponds to a 3.3 cm spatial resolution. The slow rise edge and the high dark count rate of the $6 \times 6\ mm^2$ SiPM make it difficult to improve the temporal resolution.

The energy calibration can be done by using the Compton edge of a $^{137}Cs$ spectrum. The relationship between the spatial resolution and the deposited energy of the recoil proton can be estimated using the quenching factor.

The neutron/gamma discrimination is a problem for the plastic scintillator. One possible solution is to use the coincidence with the neutron TPC. The sum of the scattering angle and the recoil angle (90 degree for proton) can be used as a criterion to select the neutron events.

## Acknowledgments

We thank Dr. Wei Shen for the useful discussion about the SiPM readout and measurements.